\documentclass[11pt]{article}

\usepackage[dvips]{graphicx}
\usepackage{amsmath,amssymb}

\def\ga{\mathrel{\mathpalette\fun >}}
\def\fun#1#2{\lower3.6pt\vbox{\baselineskip0pt\lineskip.9pt
\ialign{$\mathsurround=0pt#1\hfil ##\hfil$\crcr#2\crcr\sim\crcr}}}

  \newcommand{\be}{\begin{equation}}
\newcommand{\ee}{\end{equation}}

\newcommand{\vek}{\mbox{\boldmath${\rm k}$}}

\newcommand{\vep}{\mbox{\boldmath${\rm p}$}}

\newcommand{\vegamma}{\mbox{\boldmath${\rm \gamma}$}}

\title{\large{\textbf{QUARK DIFFUSION ALONG   THE CRITICAL LINE}}}
\begin{document}

\date{ }
\maketitle

\begin{center}
\vspace*{-1cm}
\author{ \small{B. Kerbikov}}

\small{\itshape{State Research Center\\
Institute of Theoretical and Experimental Physics,\\
Moscow, Russia}}
\end{center}

\begin{center}

\footnotesize{The quark diffusion coefficient in the critical
window near $T_c$ and at moderate values\\ of the  chemical
potential is evaluated.

\vspace{0.5cm} Contribution to the proceedings of the Rencontres
de Moriond on QCD and Hadronic\\ Interactions, La Thuile, Italy,
March 2008.}

\end{center}

\section*{\normalsize{1. Introduction}}

Evaluation of the QCD transport coefficients at $T,\mu>0$ is
relevant for the analysis of heavy ion collisions at RHIC and LHC.
We present a simple  expression for the quark diffusion
coefficient in the vicinity  of the phase transition line at
moderate values of the quark chemical potential, i.e., at $\mu\sim
300-600 $ MeV. The physics of this region has been in detail
analyzed in [1]. It is characterized by drastic increase of
fluctuations of both quark and gluon fields, the corresponding
Ginzburg-Levanyuk number is $Gi\ga 10^{-2}$, which is by about ten
orders of magnitudes larger than for metal superconductors.

The transport coefficients can be expressed in terms of the
time-dependent propagator [2,3]. In writing the expression for the
fluctuation quark propagator (FQP) use will be made of the two
assumptions. First, we assume that at such  values of $\mu$ the
quark Fermi surface is already formed and hence momentum
integration can be performed around it in the same way as in the
BCS theory of superconductivity. Second, the fluctuation regime
will be treated following the guidlines of the condensed matter
theory of disordered systems. The only relevant parameter entering
into the diffusion coefficient will be  the  collision, or
relaxation time $\tau$. This parameter may be in turn calculated
in dynamical models, e.g., in the Langevin picture. Denoting the
FQP as $L(\vep,\omega)$ we may define it  as [2,3] \be
L^{-1}(\vep,\omega) =-\frac{1}{g}+ F(\vep,\omega),\label{1}\ee \be
F(\vep,\omega) = \sum_k G^R (\vek, k_4) G^A(\vep-\vek,
\omega-k_4),\label{2}\ee Here $g$ is the coupling constant with
the dimension $m^{-2}$, the sum over $k$ implies the momentum
integration and Matsubara summation, $k_4=-\pi (2n+1)T, ~G^R$ is
the thermal retarded  Green's function  which reads \be G^R(\vek,
k_4) =i (\vegamma \vek + \gamma_4 k_4 -im + i\mu \gamma_4
-\frac{1}{2\tau})^{-1},\label{3}\ee and $G^A=(G^R)^*$. We shall
compute $F(\vep, \omega)$ in the approximation of long-wave
fluctuations \be F(\vep,\omega) \simeq A(\omega)
+B\vep^2,\label{4}\ee where the frequency dependence of  $B$ is
neglected. Finally the contribution from antiquarks is also
neglected in  the  present short contribution. The evaluation of
(\ref{2}) is straightforward though somewhat cumbersome. First we
perform momentum integration around the Fermi surface, then
Matsubara summation. The FQP exhibits the diffusion pole yielding
the following  answer for the diffusion coefficient $D$

\be D=-\frac{8T\tau^2}{3\pi} \left\{ \psi \left(
\frac12+\frac{1}{4\pi T\tau}\right) -\psi \left(\frac12\right)-
\frac{1}{4\pi T\tau} \psi'
\left(\frac12\right)\right\},\label{5}\ee where $\psi(z)$ is the
logarithmic derivative of the $\Gamma$-function. This expression
is valid both below and above $T_c$. From (\ref{5}) one obtains
two simple limiting regimes
$$ ~~~~~~~~~~~~~~~~~~~~~D\simeq\left\{\begin{array}{llr}  \tau/3,& T\tau\ll
1,&~~~~~~~~~~~~~~~~~~~~~~~~~~~~~~~~~~~~~~~~~~~~~~~~~~~~~~~~~~~(6)\\1/6T,&
T\tau\gg
1.&~~~~~~~~~~~~~~~~~~~~~~~~~~~~~~~~~~~~~~~~~~~~~~~~~~~~~~~~~~~
(7)\end{array}\right.$$Expression(6) may be called relativistic
Drude equation.

\section*{\normalsize{Acknowledgments}}

It is a pleasure to thank Tran Thanh Van for the invitation and
warm hospitality  and Bolek Pietrzyk for making my participation
possible. The support from RFBR qrants 06-02 17012, 08-02-08082
and  NSh-4961.2008.2 is gratefully acknowledged.

\section*{\normalsize{References}}

1. B. Kerbikov and E. Luschevskaya,  Phys. Atom. Nucl., {\bf 71},1
(2008).\\ 2. A. Larkin and A. Varlamov, cond-mat/0109177, in: The
Physics of Conventional and Unconventional Superconductors. Eds.
K. Bennermann,\\ J. Ketterson, Springer-Verlag, Berlin-Heidelberg,
2001.\\3. B. Kerbikov, Proc. XLIInd Recontres de Moriond, edited
by E.Aug\'{e}, B.Pietrzyk and J.Tran Thanh Van, p.283.

\end{document}